\documentclass{webofc}
\usepackage[varg]{txfonts}   
\usepackage{graphics}
\newcommand\nn{\nonumber}                  
\woctitle{QCD@Work 2014}
\begin{document}
\hfill IFJPAN-IV-2014-14
\title{Analysis of BaBar data for three meson tau decay modes using the Tauola generator}
%
%

\author{Olga Shekhovtsova\inst{1,2}\fnsep\thanks{\email{olga.shekhovtsova@ifj.edu.pl}} 
}

\institute{Institute of Nuclear Physics PAN ul. Radzikowskiego 152 31-342 Krakow, Poland
\and
           Kharkov Institute of Physics and Technology   61108, Akademicheskaya,1, Kharkov, Ukraine
          }

\abstract{%
The hadronic current for the  $\tau^- \to \pi^-\pi^+\pi^-\nu_\tau$ decay calculated in the framework of the Resonance Chiral Theory with an additional modification to include the $\sigma$ meson is described. Implementation into  the Monte Carlo generator Tauola  and fitting strategy to get the model parameters using the one-dimensional distributions are discussed. The results of the fit to  one-dimensional mass invariant spectrum of the BaBar data  are presented.
This paper is based on \cite{Nugent:2013hxa}.  

}
\maketitle
\section{Introduction}
\label{intro}
The precise experimental data for tau lepton decays collected at B-factories (both Belle and BaBar) provide an opportunity to measure the Standard Model (SM) parameters, such as the strong coupling constant, 
the quark-mixing matrix, the strange quark mass etc, and for searching new physics, beyond SM. 
The leptonic decay modes of the tau lepton allow to test the universality of the lepton couplings to the gauge bosons. The hadronic decays (in fact, the tau lepton due to its high mass is only one that 
can decay into hadrons) give an information about the hadronization mechanism and resonance dynamics in the energy region where the methods of the perturbative QCD cannot be applied. Also hadronic flavour-violating and CP violating decays of tau lepton allow to search for new physics scenario.

Hadronic tau lepton decays are also a tool in high-energy physics. At the LHC and future linear colliders a correct simulation of the hadronic decay modes, mainly two pion and three pion modes, is needed to measure the Higgs spin and its CP properties.

The implementation of the appropriate information on the hadronization of the
QCD currents represents a key task of the TAUOLA library \cite{Jadach:1993hs,Jadach:1990mz}. TAUOLA is a Monte Carlo generator (MC) dedicated to generating tau decays and it is used in the analysis of experimental data both at B-factories and LHC. It is important to include in the analyses the information of QCD itself and not of ad-hoc
models that may screen the appropriate information from data. On the other hand an agreement with experimental data is essential and verifies a theoretical model.
 Resonance Chiral
Theory (RChT) \cite{Ecker:1988te,Ecker:1989yg} provides such a reliable framework as it has been shown in many
previous publications \cite{Dumm:2009va,Dumm:2009kj,Dai:2013joa,Dubinsky:2004xv}.  A set of RChT currents for the main two meson and three meson, namely, $\pi^-\pi^0$, $K^-\pi^0$, $K^0\pi^-$, $\pi^-\pi^-\pi^+$, $\pi^0\pi^0\pi^-$, $K^-\pi^- K^+$, $K^0\pi^- \bar{K}^0$ and $K^-\pi^0 K^0$, was installed.  That set covers more than $88\%$ of total hadronic $\tau$ width.  The implementation of the currents, technical tests on it as well as necessary theoretical concepts  are documented in \cite{Shekhovtsova:2012ra}. 

Publication by BaBar collaboration of the one-dimensional distributions for the $\pi^-\pi^-\pi^+$ mode \cite{Nugent:2013ij} allows us to compare the RChT predicted spectra and modify the corresponding hadronic current to ?describe? the experimental data. The main change is related with the $\sigma$ meson inclusion.
The paper is organized as follows. In Section 2 the hadronic currents for $\tau^-\to \pi^-\pi^-\pi^+\nu_\tau$ is presented. Fit  of the three mass-invariant distributions for that process to BaBar data is presented in Section 3 where also the numerical tests are discussed. Summary, Section 4, closses the paper.

\section{Hadronic current for $\tau^- \to \pi^-\pi^-\pi^+ \nu_\tau$ mode}
\label{sect_curr}

For the final state of three pions $\pi^-(p_1)$, $\pi^-(p_2)$, $\pi^+(p_3)$ the Lorentz invariance 
determines the decomposition 
of the hadronic current to be \cite{Shekhovtsova:2012ra}
\begin{eqnarray}
J^\mu =N \bigl\{T^\mu_\nu \bigl[ (p_2-p_3)^\nu F_1  -  (p_3-p_1)^\nu
 F_2  \bigr]
+ q^\mu F_4  -{ i \over 4 \pi^2 F^2}      c_5
\epsilon^\mu_{.\ \nu\rho\sigma} p_1^\nu p_2^\rho p_3^\sigma F_5      \bigr\},
\label{fiveF}
\end{eqnarray}
where:  $T_{\mu\nu} = g_{\mu\nu} - q_\mu q_\nu/q^2$ denotes the transverse
projector, and $q^\mu=(p_1+p_2+p_3)^\mu$ is the momentum of the hadronic system.  The normalization factor is
 $N = \mathrm{cos} \theta_{\mathrm{Cabibbo}}/F$, where $F$ is the pion decay constant in chiral limit. 
In the isospin symmetry limit, the $F_5$ form factor for the three pion mode is zero due to $G$-parity conservation 
\cite{Kuhn:1992nz} 
and thus we will neglect it. 

The hadronic form factor, $F_i$, are model dependent functions. In general they depend on three independent invariant masses that are constructed from the three meson 
four-vectors. We chose $q^2=(p_1+p_2+p_3)^2$
and two invariant masses  $s_1=(p_2+p_3)^2$, $s_2=(p_1+p_3)^2$ built from 
pairs of momenta (then $s_3=(p_1+p_2)^2 = q^2-s_1-s_2+3m_\pi^2$).

In the framework of RChT  every hadronic form factor consists of three parts: a chiral contribution (direct decay,
without production of any intermediate resonance), one-resonance and double-resonance mediated processes.  The exact form of the form factors within RChT are written in \cite{Shekhovtsova:2012ra}, Eqs. (4)-(10). We would like to stress that only vector and axial-vector resonances contribution to the hadronic form factors were included \cite{Dumm:2009va,Shekhovtsova:2012ra}. The first comparison of the $R\chi L$ results for the $\pi^-\pi^-\pi^+$ mode with the 
BaBar data 
\cite{Nugent:2013ij}, did not demonstrate a satisfactory agreement for the two pion invariant mass distributions and hinted that the 
lack of the $f_0(600)$ (or $\sigma$) meson contribution to the hadronic form factors  may be responsible for that discrepancy \cite{Shekhovtsova:2013rb}.

As the $\sigma$ meson is, predominantly, a teraquarkj state it cannot be included in the RChT formalism. In view of this 
we have decided to incorporate the $\sigma$ meson following a phenomenological approach as the s-wave Breit-Wigner function. In fact, a similar parametriztaion was used by the CLEO collaboration in the analysis of the three pion decay modess of the tau lepton \cite{Shibata:2002uv}. The $\sigma$ meson inclusion affects the $F_1(Q^2,s,t)$ and  $F_2(Q^2,s,t)$ form factors in the following way%

\begin{eqnarray}\label{eq:ff_sig}
\!\!\!\!\!\!\!\!\!\!\!\!\!\!F_1^{\mbox{\tiny R}}&\rightarrow&  F_1^{\mbox{\tiny R}}+\frac{\sqrt{2}F_VG_V}{3F^2}
\left[\alpha_\sigma BW_\sigma(s_1)F_\sigma(q^2,s_1) +\beta_\sigma BW_\sigma(s_2)F_\sigma(q^2,s_2)\right]\, ,\\
\!\!\!\!\!\!\!\!\!\!\!\!\!\!F_1^{\mbox{\tiny RR}}&\rightarrow&F_1^{\mbox{\tiny RR}}+\frac{4F_AG_V}{3F^2}
\frac{q^2}{q^2-M_{a_1}^2-iM_{a_1}\Gamma_{a_1}(q^2)} 
\left[\gamma_\sigma BW_\sigma(s_1)F_\sigma(q^2,s_1)+\delta_\sigma BW_\sigma(s_2)F_\sigma(q^2,s_2)\right] ,\nonumber
\end{eqnarray}
where
\begin{eqnarray}\label{eq:bw_sig}
BW_\sigma(x) = \frac{M_\sigma^2}{M_\sigma^2 -x -iM_\sigma \Gamma_\sigma(x)} ,  \; \; \Gamma_\sigma(x) = \Gamma_\sigma\frac{\sigma_\pi(x)}{\sigma_\pi(M_\sigma^2)} ,   
\; \; 
F_\sigma(q^2,x) = \mathrm{exp}\left[\frac{-\lambda(q^2,x,m_\pi^2)R_\sigma^2}{8q^2}\right] , \nn
\end{eqnarray}
and  $\sigma_\pi(q^2) \equiv \sqrt{1 - 4 m_\pi^2/q^2}$ and  $\lambda(x,y,z) = (x - y -z)^2 - 4yz$.
Bose symmetry implies that the form factors $F_1$ and
$F_2$ are related $F_2(q^2,s_2,s_1) =  F_1(q^2,s_1,s_2)$. The vertex coupling constants $\alpha_\sigma$, $\beta_\sigma$, $\gamma_\delta$ and $\delta_\sigma$ as well as the mass ($M_\sigma$) and width ($\Gamma_\sigma$) of the $\sigma$ mesons are left to be fitted to data.  More details about the modification to the RChT three pion currsnt are presented in \cite{Nugent:2013hxa}.

The further application we present here also a result for the differential $\tau \to \pi^-\pi^-\pi^+\nu_\tau$ width:
\begin{equation}\label{eq:wid3pi}
\frac{d\Gamma}{dq^2 ds_1 ds_3} = \frac{G_F^2|V_{ud}|^2}{128(2\pi)^5 M_\tau F^2}\bigg(\frac{M_\tau^2}{q^2}-1\bigg)^2
\bigg[ W_{SA} + \frac{1}{3} \bigg(1+2\frac{q^2}{M_\tau^2}\bigg) W_A\bigg] , 
\end{equation}
where
\begin{equation}\label{eq:spect_function}
W_{A } = - (V_1^\mu F_1 +V_2^\mu F_2 + V_3^\mu F_3 )(V_{1\mu} F_1 + V_{2\mu} F_2 + V_{3\mu} F_3 ) \, ,
\; \; \; \; \; \; \; W_{SA} = q^2 |F_4|^2 \nonumber \,.
\end{equation}

\section{Fit to $\tau^- \to \pi^-\pi^-\pi^+ \nu_\tau$ data from BaBar. Numerical results and tests}\label{sect_fit}

The three one-dimensional distributions, namely $d\Gamma/dq^2$, $d\Gamma/ds_1$ and $d\Gamma/ds_3$, were fitted to the BaBar data \cite{Nugent:2013ij}. The corresponding distributions are obtained from the three-dimensional spectrum, Eq.(\ref{eq:wid3pi}), by integration over two parameters. The partial width is normalized to one measured by BaBar $\Gamma= (2.00\pm 0.03\%)\cdot 10^{-13}$ GeV \cite{Aubert:2007mh}. 

\begin{figure*}
\centering
\includegraphics[scale=.22]{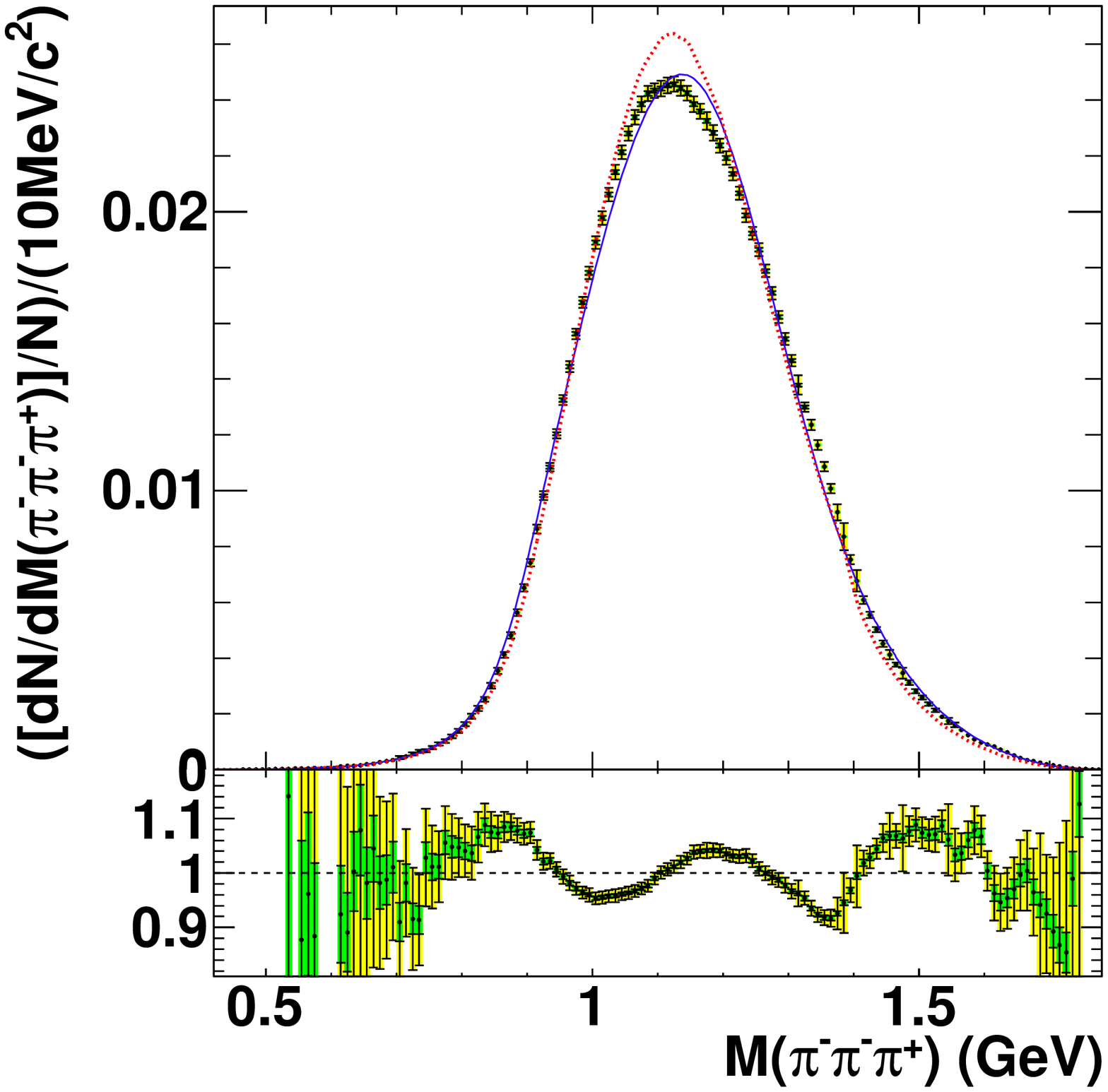}
\includegraphics[scale=.220]{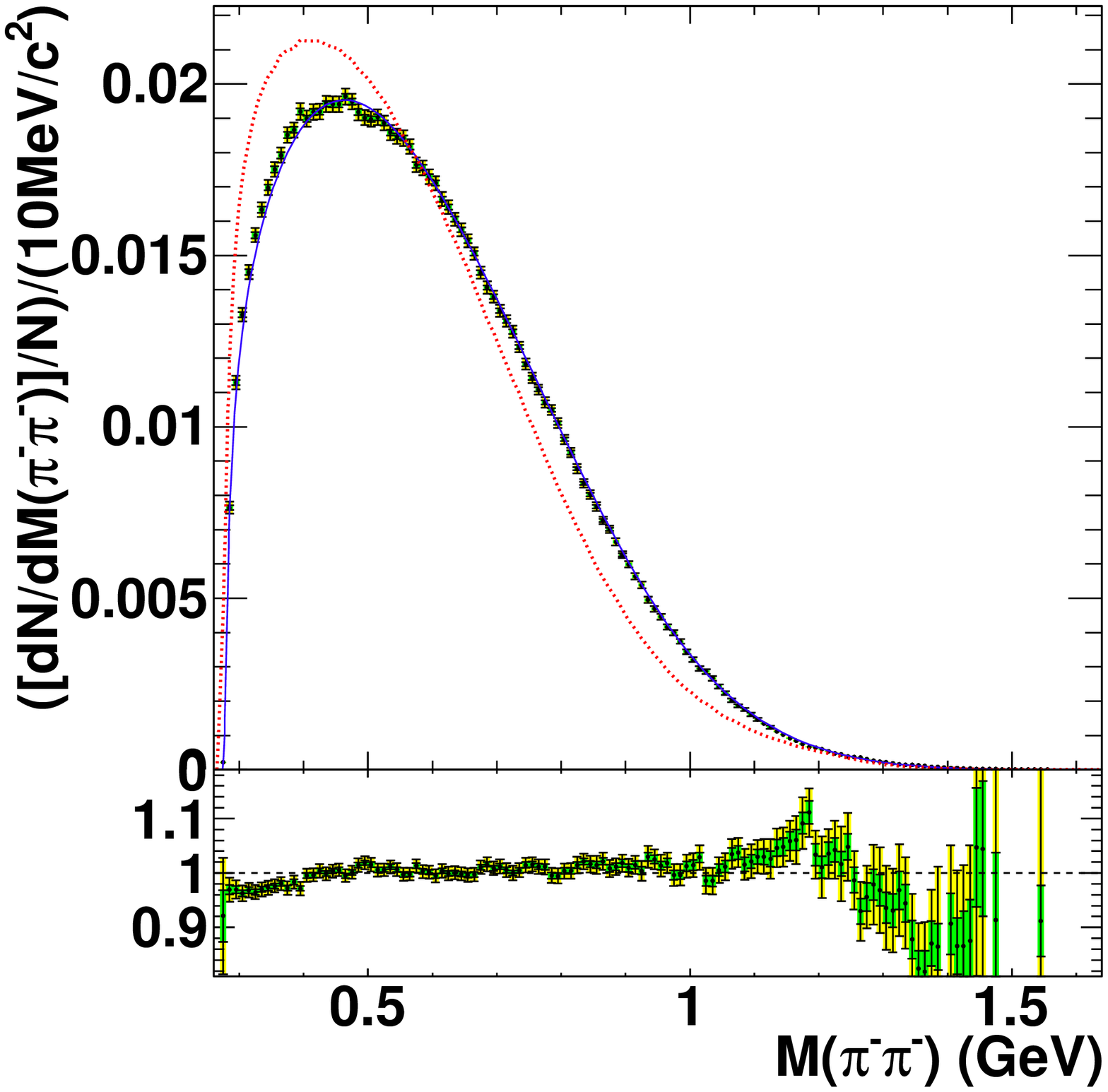}
\includegraphics[scale=.220]{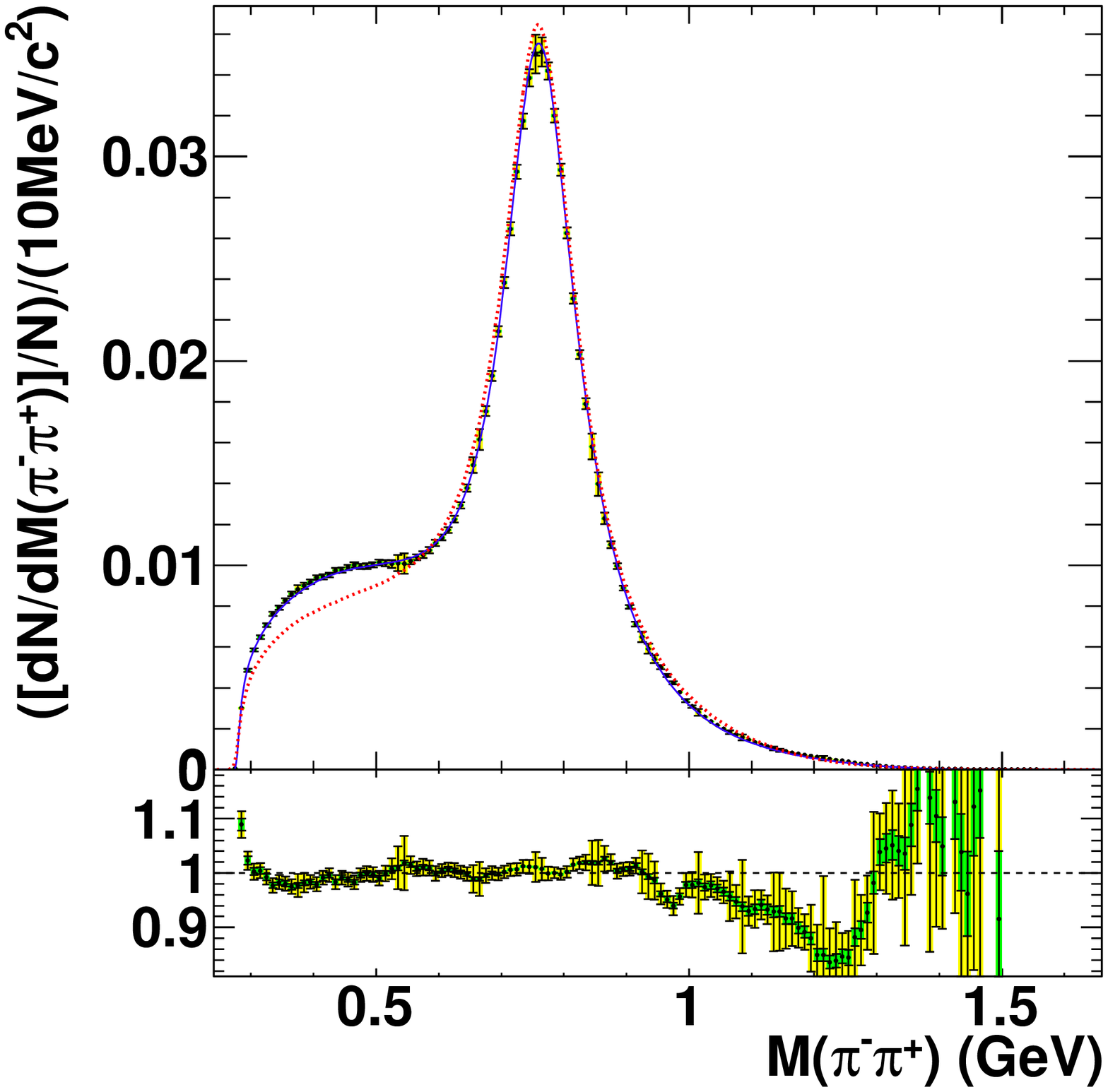}
\caption{The $\tau^- \to \pi^- \pi^-\pi^+\nu_\tau$ decay 
invariant mass distribution of the three-pion system (left panel) and two-pion pairs (central and right panels).  The BaBar measurements \cite{Nugent:2013ij} are represented 
by the data points, 
with the results from the RChT current as described in the text (blue line) and the
 old tune from CLEO from Refs.~\cite{Davidson:2010rw} (red-dashed line) overlaid.   
At the bottom of the figures ratio 
of new  RChT prediction to the data is given.
The parameters used in our new model are collected in Table  \ref{tab:fit}. 
\label{fig:res}}
\end{figure*}

The fit results are presented in table~\ref{tab:fit} and figure~\ref{fig:res} and correspond to $\chi^2/ndf = 6658/401$.  In our previous paper \cite{Shekhovtsova:2013rb}, $\chi^2$ was computed using 
the combined statistical 
and systematic uncertainties since only the total covariance matrix was publicly available. For the present results 
we obtain $\chi^2/ndf = 910/401$, when the 
total covariance matrix is used and conditions enabling direct
comparisons are fulfilled,
 that is eight times better than the 
previous result \cite{Shekhovtsova:2013rb}.
\begin{table*}
\begin{small}
\begin{tabular}{|l|l|l|l|l|l|l|l|l|}
\hline
&$M_\rho$&$M_{\rho'}$&$\Gamma_{\rho'}$&$M_{a_1}$&$M_\sigma$&$\Gamma_\sigma$&$F$&$F_V$ \\
\hline
Min&0.767&1.35&0.30&0.99&0.400&0.400&0.088&0.11\\
\hline
Max&0.780&1.50&0.50&1.25&0.550&0.700&0.094&0.25\\
\hline
Fit&0.771849&1.350000&0.448379&1.091865&0.487512&0.700000&0.091337&0.168652 \\
\hline
\end{tabular}
\begin{tabular}{|l|l|l|l|l|l|l|l|}
\hline
      & $F_{A}$ & $\beta_{\rho'}$ & $\alpha_\sigma$ & $\beta_\sigma$ & $\gamma_\sigma$ & $\delta_\sigma$ & $R_\sigma$ \\
\hline
Min   & 0.1 &  -0.37 & -10.  & -10.  &  -10.   & -10.      &  -10.     \\
\hline
Max   & 0.2 & -0.17 & 10.   & 10.  &  10.     &  10.     &  10.   \\
\hline
Fit   & 0.131425 & -0.318551 & -8.795938 & 9.763701 & 1.264263 & 0.656762 & 1.866913  \\
\hline
\end{tabular}
\end{small}
\caption{Numerical ranges of the RChT parameters
used to fit the BaBar data 
and the result of fit to BaBar data for three pion mode \cite{Nugent:2013ij} . 
}
\label{tab:fit}
\end{table*}

The statistical uncertainties were determined using the {\tt HESSE}
routine from {\tt minuit}~\cite{James:1975dr} 
under the assumption 
that the correlations between distributions and the correlations related to having two entries per event in 
the $\pi^{-}\pi^{+}$ distribution can be  neglected. The fit results with estimated systematical and statistical errors, the statistical correlation matrix and the correlation matrix for systematic uncertainties are collected in tables 3, 4  and 5 of \cite{Nugent:2013hxa}, correspondingly. The strong correlation (correlation coefficients moduli
bigger than 0.95) was found between four parameters of the model $M_{a_1}$, $F_\pi$, $F_V$ and $\beta_{\rho'}$. The correlation between these parameters can be explained by the underlying dynamics: the  dominant contribution to the hadronic currents originates
 from the exchange $a_1 \to (\rho; \rho')\pi$ and, as a consequence,  strong correlations between $F_V$, $F_A $, $F_\pi$ and also $M_{a_1}$ and 
$\beta_{\rho'}$ could have been expected, as it is the case for all of them but for 
$F_A$ which shows slightly smaller correlations (more details can be found in section V.A in \cite{Nugent:2013hxa}).  
  Also the parameters $\beta_\sigma$ and $\Gamma_{\rho'}$ are correlated (the corresponding correlation coefficients are larger tham 0.85).

The following test has been done to check whether the obtained minimum is a global one and does not depend on an initial point. We started with random scan of  $2.1*10^5$ points and select 1000 events with the best $\chi^2$, from which 20 points with maximum distance between points and then these points are used as a start point for the full fit. The result:
more than an half converges to the minimum (table~\ref{tab:fit}), others either fall with number of parameters at their limits or converge to local minimum with higher $\chi^2$. Therefore, we conclude that the obtained result is stable and does not depend on an initial value of the fitting parameters. 

As an additional cross check we calculated the partial width resulting from the phase space integration
of the matrix element $\Gamma_{\tau^- \to \pi^- \pi^-  \pi^+ \nu_\tau } = 1.9974 \cdot 10^{-13}$ GeV which
agrees with the one measured by BaBar    $\Gamma_{\tau^- \to \pi^- \pi^-  \pi^+ \nu_\tau } = (2.00\pm 0.03\%)\cdot 10^{-13}$ GeV \cite{Aubert:2007mh}.

Comparison between the RChT results (after fit) and the BaBar spectra, presented in figure~\ref{fig:res}, demonstrates a possibility of missing resonances in the model~\ref{eq:ff_sig}. For the $\pi^+\pi^-$ mass invariant spectrum it  can be f2(1270) and f0(1370), which
were reported CLEO \cite{Shibata:2002uv,Asner:1999kj} and are suggested by the fit
to the BaBar data. The largest discrepancies between data and the fitted distribution,
which are responsible for a significant part of the total $\chi^{2}$, 
are observed also in the 3$\pi$ invariant mass distribution. 
The slope and shape of the disagreement in the 3$\pi$ invariant mass spectrum, in particular around $1.5$ GeV in
Fig. \ref{fig:res}, indicates the possibility of interference between
$a_1(1260)$ and its excited state $a_1(1640)$.

\section{Conclusion}\label{sect_concl}

In this paper we discussed the hadronic current for the $\tau^- \to \pi^-\pi^+\pi^-\nu_\tau$ decay within RChT and a  modification to the current to include the sigma meson.   The choice of this channel was motivated by its
relatively large branching ratio, availability of unfolded experimental distribution and already non-trivial 
dynamics of three-pion final state.  
In addition, this channel is important for Higgs spin-parity studies through the associated di-$\tau$ decays.  As a result, we improved agreement with the data by a factor of 
about eight.

To get the numerical values of the RChT parameters we fitted the one dimentional mass invariant distributions to the published BaBar data. Also we have tested that the obtained results correspond to a global minimum and that the fitting procedure does not depend on the initial values of the model parameters.

We have found  discrepancies in the high mass region of the $\pi^{+}\pi^{-}$  and $\pi^{+}\pi^-\pi^- $invariant mass 
indicate the possibility of missing resonances in our $R\chi L$ approach. 
This is consistent with the observation of additional resonances, more
specifically the $f_{2}(1270)$ and
$f_{0}(1370)$ and  $a_1(1640)$,  by CLEO in
\cite{Shibata:2002uv,Asner:1999kj}.  Although we could add phenomenologically the contribution of these resonances to the amplitude, we prefer not to do it at the moment to keep a compromise between the number of parameters, the stability of the fit and the amount of experimental data.  Certainly, that type pf improvements will be done  in future analysis of multi-dimensional distributions.

The work on a generalization of the fitting strategy to a case of an arbitrary three meson tau decay is in progress.

\section{Acknowledgements}
This research was supported in part by Foundation of Polish Science
grant POMOST/2013-7/12, that is co-financed from European Union, Regional
Development
Fund and from funds of Polish National Science
Centre under decisions  DEC-2011/03/B/ST2/00107.

%
%

\end{document}